\pdfoutput=1
\documentclass[twocolumn,amssymb,floatfix,deluxe,prd,showkeys]{revtex4-1}
\usepackage{amsmath}
\usepackage{graphicx}% Include figure files
\usepackage{dcolumn}% Align table columns on decimal point
\usepackage{bm}% bold math
\usepackage{epsfig}

\begin{document}

%\preprint{APS/123-Q}
\title{The calorimetric spectrum of the electron-capture decay of $^{163}$Ho.\\
A preliminary analysis of the preliminary data}
% Force line breaks with \\

%\author{Author1}%
%\affiliation{here \\
%There}
%

\author{A. De R\'ujula}
\affiliation{IFT(UAM), Madrid, Spain; CERN,
1211 Geneva 23, Switzerland}%
\author{M. Lusignoli}
\affiliation{Sapienza, Universit\`a di Roma, and INFN, Sezione di Roma
Piazza A. Moro 2, I-00185 Roma, Italy}

\date{\today}% It is always \today, today,

\begin{abstract}
It is in principle possible to measure directly
the electron neutrino mass (or masses and mixing angles)
in weak electron-capture decays. The optimal nuclide in this respect is $^{163}$Ho.  
The favoured experimental technique, currently pursued in
 various experiments (ECHo, HOLMES and NuMECS) is ``calorimetric''.
 The calorimetric energy spectrum is a sum over the unstable vacant orbitals, or ``holes'',
 left by the electrons weakly captured by the nucleus.
 We discuss the current progress in this field and analize the
 preliminary data. Our conclusion is that, as pointed out by Robertson,
 the contribution of two-hole states is not negligible. But --in
 strong contradistinction with the tacit conclusion of
 previous comparisons of theory and observations--
 we find a quite satisfactory agreement. A crucial point is that,
 in the creation of secondary holes, electron shakeoff and not only 
 electron shakeup must be taken into account.

%After an initial surge of measurements,
%the EC approach did not seem to be competitive.
%But very recently, there has been great progress on micro-calorimeters
%and the measurement of atomic mass
%differences. Meanwhile, the beta-decay neutrino-mass limits have improved
%by a factor of 15, and the difficulty of the experiments by the cube of that figure.
%Can the ``calorimetric"  EC theory 
%cope with this increased challenge? I answer this question affirmatively.
%In so doing I briefly review the subject and extensively address some persistent
%misunderstandings of the underlying quantum physics.

\end{abstract}

%\pacs{Pacs numbers}  % PACS, the Physics and Astronomy
%Use showkeys class option if keyword
\maketitle

\section{Introduction}

In 1933 Perrin qualitatively described \cite{Perrin} and Fermi computed
\cite{Fermi} how a nonzero neutrino mass would affect the endpoint of the electron
spectrum in a $\beta$-decay process. Decades later, the laboratory quest
for a non-zero result in this kind of measurement continues in ernest \cite{Weinheimer}.

Weak electron capture (EC) has a sensitivity to the neutrino mass entirely analogous 
to the one of $\beta$-decay. EC is
the $e\,p\!\to\! \nu\,n$ weak-interaction process whereby an atomic electron interacts with a 
nucleus of charge $Z$ to produce a neutrino, leaving behind a nucleus of charge $Z-1$
and a hole in the orbital of the daughter atom from which the electron was captured.

The optimal nuclide for EC experiments is $^{163}$Ho and the most promising
technique is ``calorimetry'' \cite{ADR}, the measurement of all the energy released in 
a decay, but that of the escaping neutrino. The detailed theory of  calorimetric
energy spectra was developed in \cite{ADRML}.

The experiments ECHo \cite{ECHo}, HOLMES \cite{HOLMES} and NuMECS \cite{NUMECS}
are starting to measure the calorimetric-energy spectrum resulting from the EC
decay 
\begin{eqnarray}
&&\rm ^{163}Ho\to \, ^{163}Dy[H]+\nu_e, \nonumber\\
&&{\rm Dy[H]\to  Dy} + E_c, 
\label{decays}
\end{eqnarray}
where Dy[H] is the daughter atom, with H labelling the various
possible ``holes'' left by electrons captured from different levels of the Ho atom. 
Dy is the atom's ground state and $E_c$ is the energy released in the de-excitation of Dy[H] to Dy.
The $Q$-value for this EC decay is the record-low of the periodic table and has recently
been measured \cite{Qval} to be 
$Q\equiv M({\rm ^{163}Ho})-M({\rm ^{163}Dy})=2833\,\rm (30_{stat})\,(15_{sys})$  eV.

In the theoretical predictions of  \cite{ADRML} the calorimetric spectrum
was approximated as a sum of single-hole contributions. Robertson has pointed out
that two-hole contributions should not be negligible \cite{Robertson1,Robertson2}. Indeed,
in a EC event, the wave functions of the spectator electrons in the mother and
daugther atoms are slighly mismatched, leading to a spillover effect: secondary holes.

Faessler and his collaborators have published state-of-the art results on 
the probabilities of making two (or even three) holes in the EC decay of $^{163}$Ho
\cite{Faessler1,Faessler2}. Their calculations are relativistic and fully anti-symmetrized over the ensemble of all atomic electrons. They also employ the wave functions of the correct daughter atom, with its primary hole. 

An electron having vacated an orbital in which it leaves a second hole may have
been {\it shaken-up} to an unoccupied bound-state level of the daughter atom; or
{\it shaken-off} into the unbound ``continuum''. 

In plotting predicted spectra for $^{163}$Ho-decay calorimetry, previous authors 
\cite{Robertson2,Faessler2}
 have tacitly assumed that the computed probabilities for making a second hole characterize 
 the odds of
 electron shake-up (corresponding to 
a narrow resonance in the spectrum) and {\it not} the ones of electron shake-off  (corresponding
to a much broader feature). Our preliminary analysis of the preliminary data indicates
that this tacit assumption is entirely wrong.

\section{One- and Two-hole processes}

In the dominant branches of Ho
EC decay, Eq.\,(\ref{decays}), 
a single hole H is left in the orbitals H = M1, M2, N1, N2, O1, O2 or P1, above which $\rm Ho$ runs out of electrons. To the extent that the decay line-width of $\rm Dy[H]$ is small, this two-body decay process is ``monochromatic'', with a fixed neutrino energy for each given H, 
%The  $Q$-value is the mass difference,
% $Q=M({\rm Ho})-M({\rm Dy})\simeq 2800$ eV, 
% of these particular isotopes \cite{F2}.
 $E_\nu[{\rm H}]=Q-E[\rm H]$, where $E[\rm H]$ is the binding energy
 of the missing electron in Dy, in the convention in which it is positive.
 
 In an ideal calorimetric experiment, only the neutrino escapes the
 source-implanted detector, and the entire energy $E[\rm H]$ --delivered 
 as the electronic hole is filled-- is measured \cite{ADR,ADRML}. In the case at 
 hand the fluorescence yields
 (relative X-ray emissions) are small, the unstable Dy atom preferentially stabilizes by successive
 Coster-Kronig (and subdominant Auger) electron ejections.
 
 The spectrum of calorimetric energies, $E_c$, is a sum of Breit-Wigner
 peaks at the $E_{\rm H}$ positions with their natural hole widths, $\Gamma_{\rm H}$.
 The peak intensities are proportional to $\varphi_{\rm H}^2(0)$, the values in Ho of the
 squared wave functions at the origin 
 of the electrons to be captured. The individual contribution of a given hole to the 
 EC decay rate $R$ at a given $E_c$ is: 
 \begin{eqnarray}
&& {dR[{\rm H}]\over dE_c}= \kappa\; E_\nu\,p_\nu\,n_{\rm H}\,
\varphi_{\rm H}^2(0)\, BW[E_c,E_{\rm H},\Gamma_{\rm H}],
\label{dRdEc}\\
&& BW[E_c,E_{\rm H},\Gamma_{\rm H}]\equiv
{\Gamma_{\rm H}\over 2\pi}\,{1\over (E_c-E_{\rm H})^2+\Gamma_{\rm H}^2/4}\, ,
\label{BW}\\
&&E_\nu = (Q-E_c), \;\;\;\; p_\nu = \sqrt{(Q-E_c)^2-m_\nu^2}\, ,
\label{Eandp}
\end{eqnarray}
The factor $E_\nu$ in Eq.\,(\ref{dRdEc}) originates from the (squared) weak-interaction matrix element
and the factor $p_\nu$ in the decay's phase space. Having made explicit the $E_\nu$ factor,
$\kappa$ --in the excellent approximation in which nuclear recoil is neglected-- is a constant:
\begin{equation}
\kappa\;E_\nu \equiv {G_F^2\over 4\pi^2}\,\cos^2\theta_C\,B_{\rm H}\, |{\cal M}|^2
\label{eq:resonances}
\end{equation}
with ${\cal M}$ the nuclear matrix element, $B_{\rm H}-1$
\cite{Bambynek}  an ${\cal O}(10\%)$ correction for atomic exchange and overlap
and
$n_{\rm H}$ the electron occupancy in the H shell of Ho 
(the actual fraction of the maximum number of electrons with the quantum numbers of H).

We know from the observations of neutrino oscillations that the electron
neutrino is, to a good approximation, a superposition of three mass 
eigenstates, $\nu_i$: $\nu_e=\sum_i U_{ei} \nu_i$, 
with $\sum_i |U_{ei}|^2= 1$. Thus, we ought to have written $dR[{\rm H}]/dE_c$
in Eqs.\,(\ref{dRdEc}-\ref{eq:resonances}) as an incoherent superposition of spectra with weights $|U_{ei}|^2$
and masses $m(\nu_i)$. But the measured differences 
$m^2(\nu_i)-m^2(\nu_j)$ are so small that current direct attempts to measure
 $m_{\nu}$ are certain to reach the required accuracy
only if neutrinos are nearly degenerate in mass, in which case $m_\nu$ in
Eqs.\,(\ref{dRdEc}-\ref{eq:resonances}) stands for their nearly common mass.

%
%The contributions to the sum in Eq.\,(\ref{eq:resonances})
%have a common endpoint at $E_c=Q-m_\nu$ for all H \cite{mix}, with
%${\cal K}$ (describing the spectrum near to its endpoint), a constant
%to a level of precision sufficient for the experiments to be sensitive
%to neutrino masses of the order of a fraction on an eV \cite{ADRML,LV},
%as recently reviewed in \cite{ADR2}. This is the statement that Robertson
%contends \cite{Robertson}.
% The peaks have tails,
% so that $E_c$ extends all the way to the limit imposed by energy conservation:
% $E_c\leq Q-m_\nu$.

\subsection{Shake-up}

As Robertson pointed out \cite{Robertson1,Robertson2}, in an EC event leading to a primary hole H
there is a small probability, $P\rm (H,H')$,
for a second hole H' to be made in a shake-up process. When the second electron is shaken-up
to any unoccupied daughter-atom bound-state level --of binding energy negligible relative
to $E_{\rm Tot}=E({\rm H})+E({\rm H'})$-- the calorimetric energy peaks very close to
(but not precisely) $E_{\rm Tot}$. If this
energy is not very close to that of a single-hole peak, there is an observable feature in the
spectrum, even if $P\rm (H,H')\ll 1$.

 To the extent that the
 presence of one hole does not significantly affect the filling --i.e.~decay-- of the other,
 the natural width of a two-hole state is the sum of the partial widths:
$\Gamma_{\rm Tot}=\Gamma({\rm H})+\Gamma({\rm H'})$. In analogy
to the one-hole result of Eqs.\,(\ref{dRdEc}-\ref{eq:resonances}), the contribution
of a particular two-hole state to the calorimetric spectrum is:
\begin{eqnarray}
{dR[{\rm H,H'}]\over dE_c}&=&\kappa\; E_\nu\,p_\nu\,n_{\rm H}\,n_{\rm H'}\,
\{1-\Pi({\rm H,H')}\}
\nonumber\\
&\times&
\varphi_{\rm H}^2(0)\,P\rm (H,H')\, BW[E_c,E_{\rm Tot},\Gamma_{\rm Tot}],
\label{TwoHolesup}
\end{eqnarray}
where $n_{\rm H'}$ is the occupancy in the H' shell
and $\Pi({\rm H,H'})$ is the operator interchanging the two implicated electrons.
More precisely, $1-\Pi({\rm H,H'})$ stands for the operation of symmetrizing the
electron pair's orbital wave functions in the singlet antisymmetric spin state,
antisymmetrizing the wave functions in the triplet symmetric spin state and
adding the results with weights 1/4 and 3/4.

\subsection{Shake-off}

The creation of a second hole H' in the capture leaving a hole H can also
occur as the shake-off of the electron in the orbital H' to the ``continuum''
of unbound electrons: 
\begin{equation}
{\rm Ho}\rightarrow {\rm Dy[{\rm H,H'}]}+e^-+\nu_e.
\label{eq:2holes}
\end{equation}
In such a 3-body decay, neither the electron nor the neutrino are
approximately monochromatic. 
The neutrino energy, $E_\nu$, and the ejected electron's kinetic energy, $T_e$, satisfy
$E_\nu+T_e=Q-E[\rm Tot]$. The electron's energy and the daughter Dy ion 
energy excess add up to the observable calorimetric energy $E_c=T_e+E[\rm Tot]$.

Let $|\rm Ho[H]\rangle$ be the wave function, in Ho, of the orbital whose electron is
to be captured and $\rm {| Dy[H,H'};p_e]\rangle$ the continuum wave function
of the electron ejected  off the daughter two-hole Dy ion. 
In the sudden approximation
the shakeoff  distribution in electron momentum
$p_e$ (or in its energy $T_e$) ensues from the square of the wave function overlap:
\begin{eqnarray}
\!\!\!\!\!\!{dM\over dp_e}&=&|\{1-\Pi({\rm H,H')}\}\,\varphi_{\rm H}(0)\,
 \langle \rm {Ho[H'] | Dy[H,H'};p_e]\rangle|^2,
\label{overlapfreep}\\
\!\!\!\!\!\!{dM\over dT_e}&=&{m_e\over p_e}\,{dM\over dp_e}.
\label{overlapfree}
\end{eqnarray}

It is simplest to discuss the rate for the shake-off process of Eq.\,(\ref{eq:2holes})
by doing it for starters in the vanishing-width approximation for the daughter holes.
In this case
\begin{equation}
{dR\over dT_e}=\kappa\; E_\nu\,p_\nu\,n_{\rm H}\,n_{\rm H'}\,{p_e\over 4\pi^2}\,{dM\over dT_e}.
\label{edistr}
\end{equation}
%{\bf NOTE Ho messo tutti i $\pi$ dello spazio delle fasi qui, probabilmente mi sono
%gia preso certi che vengono ripetuti nel capitolo III. Da vedere}
The resulting $E_c$ distribution is:
\begin{equation}
{dR\over dE_c}=\int_0^{Q-E_{\rm Tot}}{dR\over dT_e}\,\delta(E_c-E_{\rm Tot}-T_e)\,dT_e
\label{Ecdistr}
\end{equation}
To undo the zero-width approximation, substitute the above $\delta$ function by
$BW[E_c-T_e,E_{\rm Tot},\Gamma_{\rm Tot}]$, with $BW$ defined as in Eq.\,(\ref{BW}).

The state-of-the-art way to estimate the size and shape of a shake-off contribution
to the calorimetric spectrum would be to extend the methods used in \cite{Faessler1} 
to unbound-electron wave functions to obtain the results of Eq.\,(\ref{overlapfree}), to
be input in Eq.\,(\ref{Ecdistr}) and ``widened'' as stated with the holes' natural widths
(and the experimental resolution function). Rather than attempting to perform
a state-of-the-art calculation, we
shall in the next section --once the data inform us of the need to do so-- make a
simple estimate for the case of N1 capture with O1 shake-off.

\section{Preliminary analysis of the preliminary data}

\begin{figure}[htbp]
\begin{center}
\hspace{.99cm}
{\includegraphics[width=.42\textwidth]{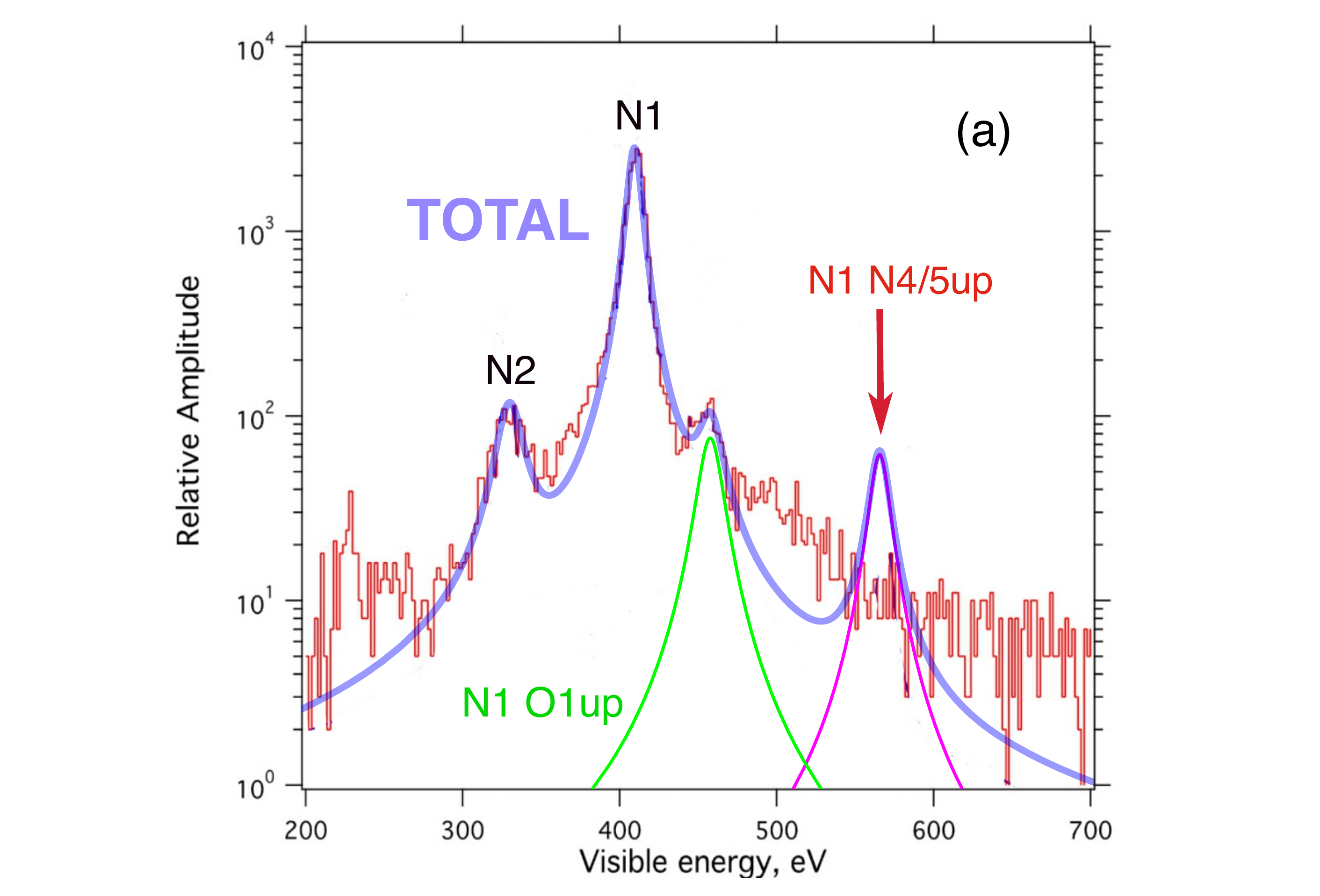}
\\
\vspace{-2cm}
\hspace{-1.2cm}
\includegraphics[width=.55\textwidth]{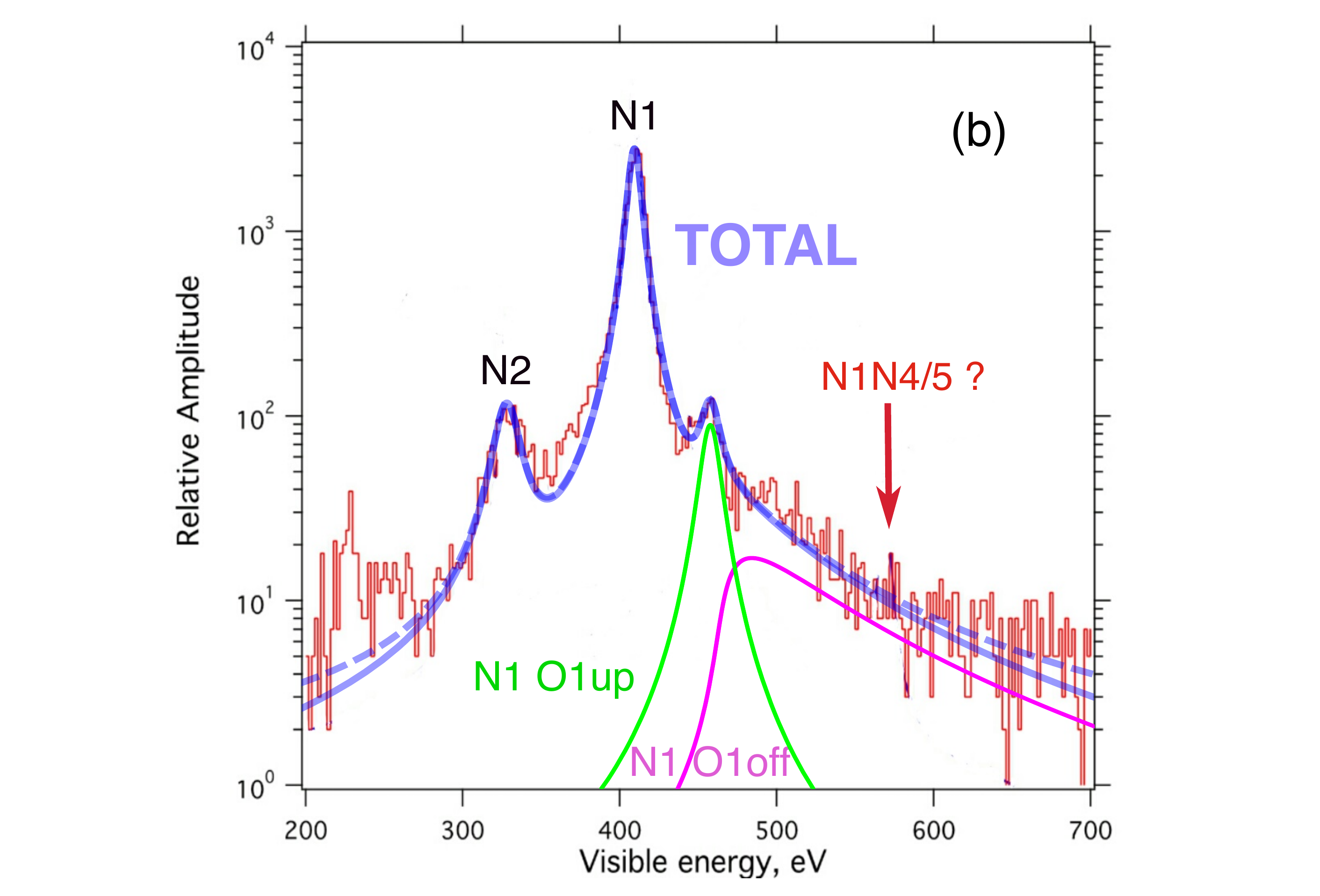}}
\end{center}
\caption{The N region of the calorimetric spectrum. 
 ECHo data \cite{ECHo}, as presented in \cite{Robertson1}.
Theroretical description with two single holes (N1, N2). 
{\bf (a)} with the addition of two double holes: N1 plus O1 shakeup and
N1 plus N4 and N5 shakeup. 
{\bf (b)} with the addition of N1 plus O1 shakeup and shakeoff
(and no N1 plus N4/5). In the dashed curve a constant background of one per bin
is assumed.
\label{fig:N1}}
\end{figure}

\begin{figure}[htbp]
\begin{center}
\hspace{-1.25cm}
\includegraphics[width=0.55\textwidth]{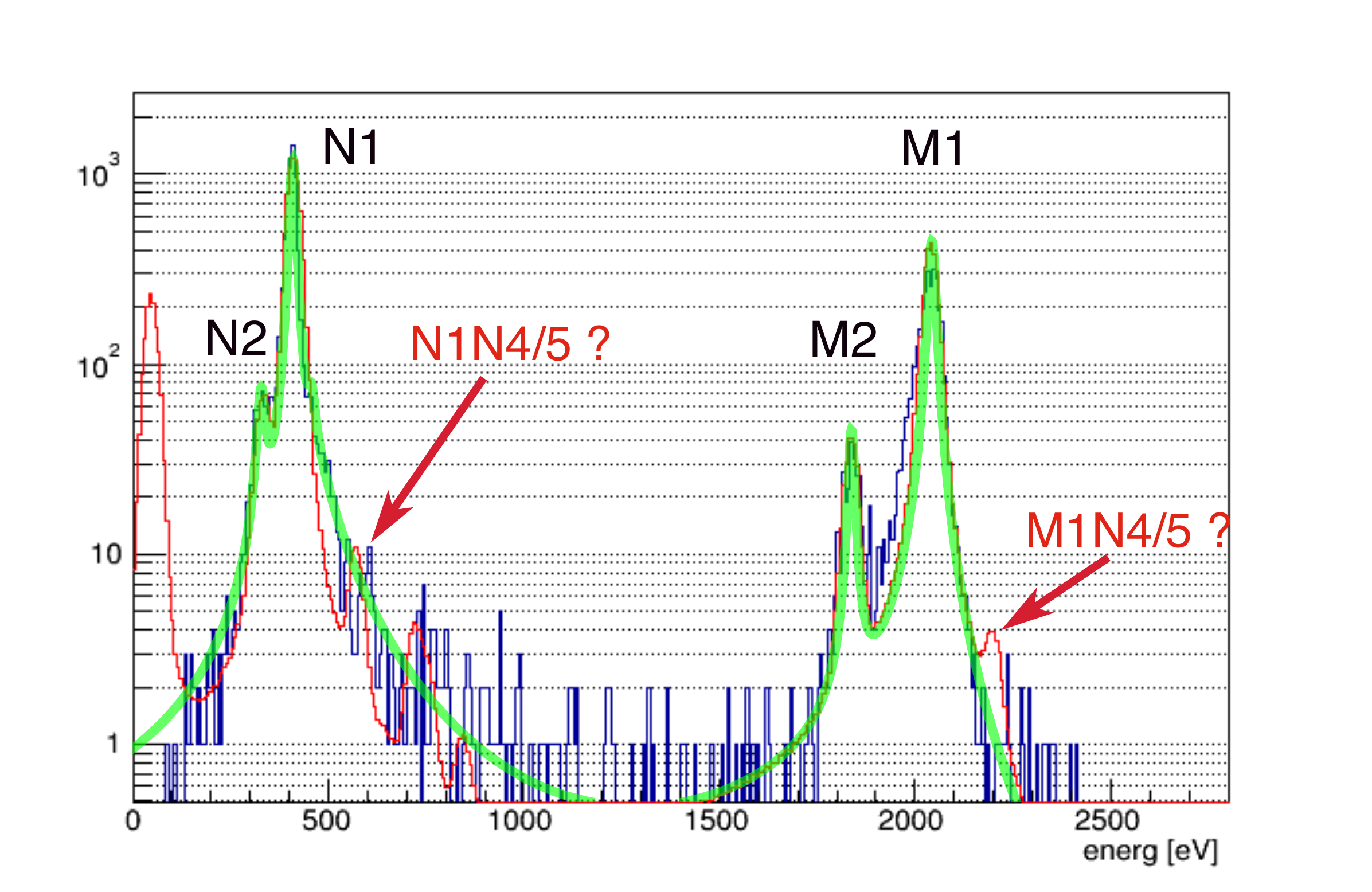}
\end{center}
\caption{Blue: the calorimetric spectrum measured by NuMECS \cite{NUMECS}.
Red: the theoretical prediction of Faessler et al. \cite{Faessler2}.
Green: the same description as in Fig.\,(\ref{fig:N1}b).
}
\label{fig:Kunde}
\end{figure}

Some recent results of calorimetric spectra in the decay of $^{163}$Ho have been
made available. We shall specifically refer to the data in the two plots of 
Fig.\,(\ref{fig:N1}) and the plot of Fig.\,(\ref{fig:Kunde}). The first data set is 
the one presented in \cite{Robertson1} based on results by members of the ECHo 
collaboration \cite{ECHo}. The data set of Fig.\,(\ref{fig:Kunde})
is from the NuMECS collaboration \cite{NUMECS}, first presented by Kunde
in \cite{Kunde}.

We shall attempt to reach a preliminary understanding of these preliminary data.
By ``understanding'' we do not mean a best fit of the theory to the data, it may be too
early for that. Relative to the theoretical expectations, we shall make small modifications
of the energies, widths and wave functions at the origin of the one-hole contributions,
so that the theoretical curves best describe --by eye-- the data for these dominant contributions
(N2, N1, M2 and M1). Similarly, we modify the
 probabilities of the two-hole contributions most relevant to this discussion
(N1O1 and N1N4/5) to provide the best description of the data. For the
single-hole contributions the
only large deviations from the theoretical values in the literature \cite{Olddata} are 
our choices of widths $\Gamma(\rm N2)=16$ eV and $\Gamma(\rm O1)=6$ eV, this last 
value being very close to the recent observation by the ECHo collaboration \cite{ECHoO1}.

 Our Fig.\,(\ref{fig:N1}a) is
somewhat similar to Fig.\,(2) of \cite{Robertson1} 
 or Fig.\,(4) of \cite{Faessler2}.
 In Fig.\,(\ref{fig:N1}a) the empirically chosen O1 normalization successfully describes the data.
The chosen N1N4/5 peak's normalization --or any other
choice-- does not succeed.
 Thus we fail to describe the data
 in the interval $\sim \! 480 \,{\rm eV}<E_c<700\, \rm eV$ with the inclusion of the
 only two-hole contribution expected 
 to be large in this domain: N1 plus N4/5 shake-up \cite{Robertson2,Faessler2}. 
 
 In Fig.\,(\ref{fig:N1}b) we have neglected the (unseen) contribution
 of N1 capture accompanied by N4/5 shakeup. More importantly, we 
 consider N1 capture with a second O1 hole in its two possible outcomes:
 that the O1 electron is shaken-up to a higher bound state, or shaken-off
 into the continuum. Once again, their respective normalizations will be fit by eye 
 to the data. 
 
 The shape of the O1 shake-up contribution in 
both panels of Fig.\,(\ref{fig:N1}) is simply
 a Breit-Wigner with a peak at $E_c=\rm E(N1)+E(O1)=464$ eV, with a natural width 
 $\Gamma({\rm N1})+\Gamma({\rm O1})=11.4$ eV, to which
 we have added in quadrature, as for all other contributions,
 an experimental resolution-width $\Gamma_{\rm exp}=8.4$ eV.
 The shape of the O1 shake-off contribution requires a longer explanation.
 
\subsection{The shake-off shape}
 
 Electron capture in Ho results in a Dy atom --which we shall
 in what follows denote as Dy*-- with a hole in the orbital
 from which the capture took place. The absent-electron
 charge partially shields the one of the absent-proton. Relative to a process without a similar
 effect --such as the creation of a primary hole by photo-ionization-- the partial shielding generally leads
 to a reduced probability for the creation of a second hole. This is because
 the wave functions of the
potentially vacated second orbitals in the parent and daughter atoms have a closer 
overlap in the presence of shielding. And --in the sudden approximation traditionally
used to make these kind of estimates-- the square of this overlap is the 
probability of creating a second hole.
 
Intemann and Pollock \cite{I&P} were the first to show how to properly treat the 
shake-off of a second electron in EC. In what follows we 
apply their method and concentrate on N1 capture accompanied by O1 shakeoff.
The trick is to treat the result of N1 capture (the absent proton and the absent electron)
 as a perturbation of the Coulomb potential of the form:
\begin{equation}
\alpha \, b(r)\equiv \alpha \left(
{1 \over r} - \int d^3r_1 \;{\vert\phi_{_{\rm N1}}(r_1)\vert^2\over \vert \vec r - \vec r_1\vert}\right).
\label{IP}
\end{equation}
 To first order in  $\alpha$ the wave function of the
O1 level in Dy* is then expressed as a linear combination of Ho eigenfunctions: 
\begin{eqnarray}
|{\rm Dy^*[O1]}\rangle\simeq |{\rm Ho[O1]}\rangle+\sum_n B_n\,|{\rm Ho}[n]\rangle
\qquad\qquad\nonumber\\
+\frac{1}{2\,\pi}\int_0^\infty dp_e \; B_{\rm off}(p_e)\; |{\rm Ho}[p_e]\rangle,\qquad\qquad 
\label{wave}\\
B_n\equiv {\alpha \over E_{_{\rm O1}}-E_n}\int d^3r\,
\phi_n^* (r)\phi_{_{\rm O1}}(r)\, b(r),\qquad\qquad 
\label{B}\\
B_{\rm off}(p_e)\equiv{\alpha \over E_{_{\rm O1}}+T_e} \int d^3r\,\phi^* (p_e,r)\phi_{_{\rm O1}}(r)\, b(r)\;\quad
\label{BB}
\end{eqnarray}
where $n$ in the wave functions $|{\rm Ho[n]}\rangle$ stands for the $l=0$ bound levels with $n \neq 5$ and (positive) binding energy $E_n$, $\phi (p_e,r)$ is the unbound wave function, defined as in \cite{L&L}, of an $l=0$ electron with momentum $p_e$ and kinetic energy $T_e=p_e^2/(2\,m_e)$.

We shall continue to refer to the creation of an \{N1,O1\} pair of holes with the 
shake-off of one electron as ``O1 shakeoff''. This in spite of the fact that, to
 take into account Fermi statistics, one must --it goes without saying--
 also do the calculations encapsulated in 
Eqs.\,(\ref{IP}-\ref{BB}) with the exchange O1 $\leftrightarrow$ N1, since the two-hole final state,
at a given $T_e$, is the same independently of which electron is captured or ejected. 
Thus, in the sudden approximation, the
square of the (monopolar) matrix element
for an electron being shaken-off with energy $T_e$ results in:
\begin{equation}
{dM\over dT_e}= {m_e\over 4\,\pi^2\,p_e}\,|\{1-\Pi({\rm N1,O1)}\}\,\varphi_{\rm N1}(0)
B_{\rm off}(p_e)|^2
\label{antiSym}
\end{equation}
for the distribution function of Eq.\,(\ref{overlapfree}).

A crucial point in estimating wave-function overlaps is to choose them with the correct
spatial scale. To provide an estimate of the shape of $dM/dT_e$ we shall use non-relativistic Coulomb 
wave functions of Ho with effective values of $Z$ chosen to reproduce the relevant
orbits' mean radii, as calculated with more precise Hartree-Fock methods \cite{McLean}.
Let $r_B\equiv 1/(\alpha \, m_e)$ denote the Bohr radius. For 
$\langle r({\rm N1})\rangle=0.555\,r_B$ and $\langle r({\rm O1})\rangle=1.420\,r_B$
the effective charges are $Z_{\rm eff}({\rm N1})=43.2$, to be used in Eq.\,(\ref{IP}) and
$Z_{\rm eff}({\rm O1})=24.5$, to be used for the bound and free wave functions
in Eq.\,(\ref{BB}).

\subsection{Back to the data}

The result of the above exercise 
--convoluted with a Breit-Wigner function
including the widths of the two holes and, in quadrature, the experimental resolution--
is shown as a magenta line in Fig.\,(\ref{fig:N1}b). Its normalization is chosen to
best accomodate the data. The ratio of O1 shake-off to O1 shake-up (the areas
under the corresponding curves, extended up to $E_c=Q$) is $\sim 1.16$.

The agreement between theory and ECHo data in Fig.\,(\ref{fig:N1}b) is satisfactory.
Above the N1 peak the description of the data is very good, excellent if we choose to
add a constant background of one per bin (dashed blue curve).
The small observed excess in the lower energy side of the N1 peak must be due
to an asymmetry of the resolution function. It may be more difficult to use this as an
excuse for the data excess below 300 eV.

In Fig.\,(\ref{fig:Kunde}) we compare the theory with the NuMECS data. The
theoretical parameters (resonance energies, widths and relative magnitudes) 
are identical to the ones in the preceeding description of the ECHo data.
Only the overall normalization and the experimental resolution have been
changed: $\Gamma_{\rm exp}=16$ eV for NuMECS.

In the N1 region of Fig.\,(\ref{fig:Kunde}), once again, the data are
better described with O1 shake-off and no N4/5 shake-up than
with the opposite pair of choices. The description of these preliminary
data is once again very satisfactory and requires the presence of the
``shoulder'' due to O1 shake-off.
The absence at the theoretically expected level
of a peak corresponding to M1 capture plus N4/5 shakeup is
also remarkable. 

\section{Theory versus fits}

\begin{figure}[htbp]
\begin{center}
%\hspace{.99cm}
{\includegraphics[width=.55\textwidth]{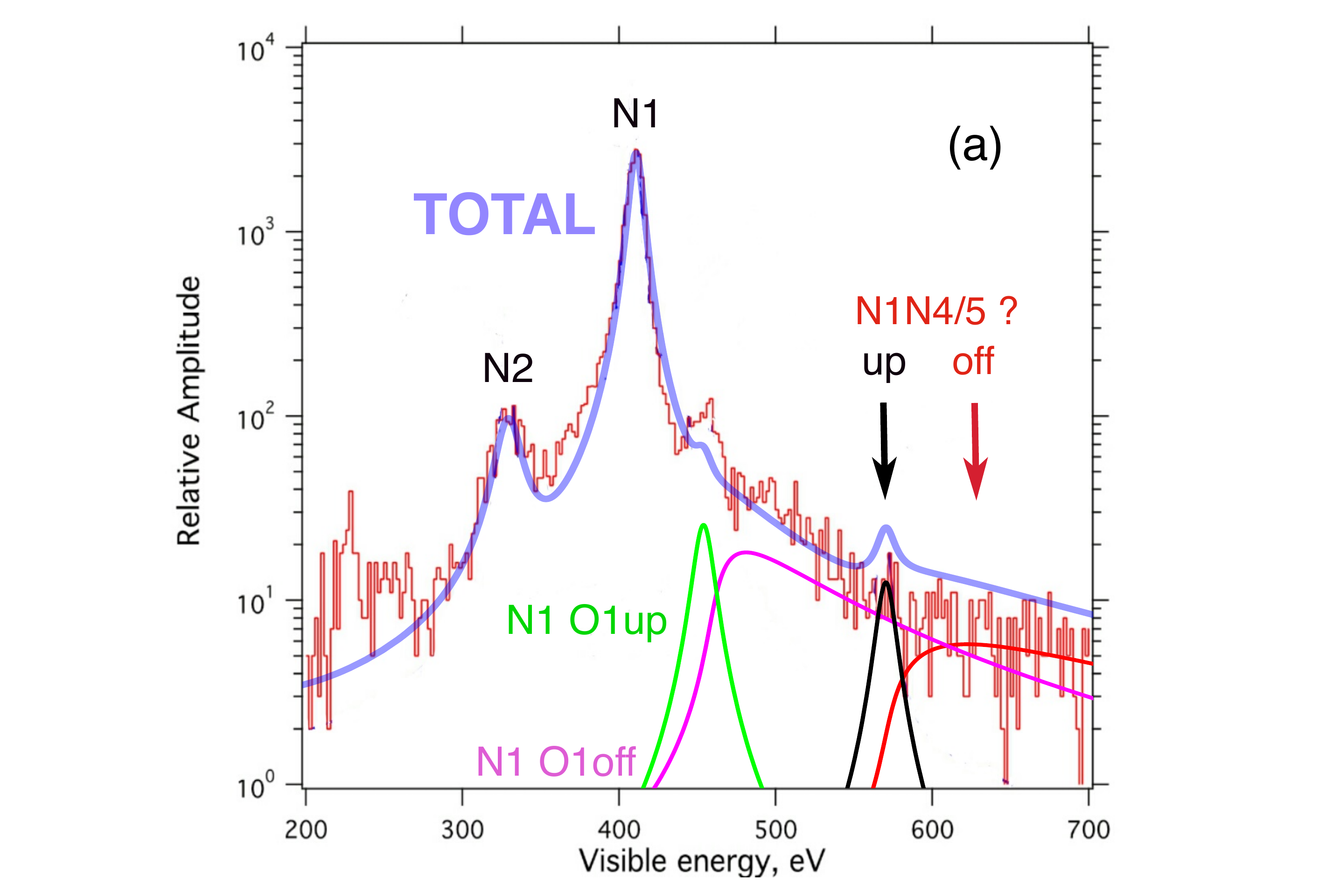}
\\
%\hspace{-1.2cm}
\includegraphics[width=.55\textwidth]{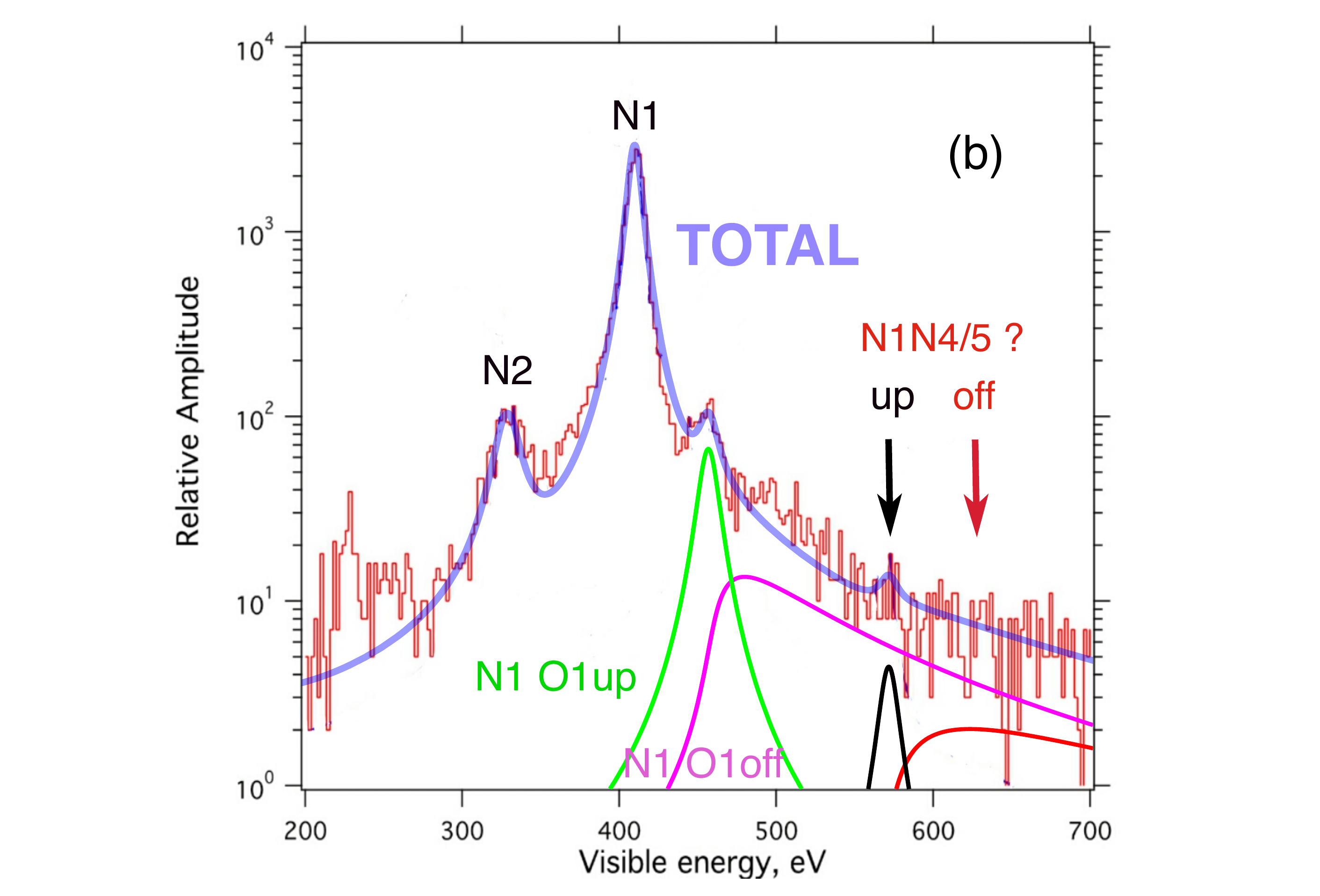}}
\end{center}
\caption{The N region of the calorimetric spectrum. 
{\bf (a)} Our simplistic theoretical prediction. {\bf (b)} With moderate modifications of the
absolute magnitudes of the various two-hole contributions. For details, see the text.
\label{fig:N1th}}
\end{figure}

In Fig.\,(\ref{fig:N1}b) we have used the predicted {\it shapes} and positions of the N1O1 
shakeup and shakeoff contributions to provide a description of the data. An important 
question concerns the extent to which the theoretical predictions for the {\it size} of 
the various two-hole contributions agree with the observations, the subject of Figs.\,(\ref{fig:N1th}a,\ref{fig:N1th}b).

The two-hole contributions expected to be the largest above the N1 peak and below 
$E_c=700$ eV are N1O1 and N1N4/5. The N1O1 shakeup contribution has a
Breit-Wigner shape and its magnitude (relative to that of the N1 peak) follows
from the sum $\sum_7^\infty B_n^2$ of Eq.\,(\ref{B}), antisymmetrized in analogy
with Eq.\,(\ref{antiSym}), in which we use the ratio $\varphi_{\rm O1}(0)/\varphi_{\rm N1}(0)$
quoted in \cite{Faessler1}. The shape and magnitude of the N1O1 shakeoff contribution are 
similarly governed by Eqs.\,(\ref{edistr},\ref{antiSym}). The N1N4/5 shakeup and shakeoff
are calculated in an identical fashion, with the substitution of O1 for N4/5, and recalling
that, neglecting nuclear-size effects, $\varphi_{\rm N4/5}(0)/\varphi_{\rm N1}(0)=0$.

In Fig.\,(\ref{fig:N1th}a) we have plotted the above-mentioned results of
our rather simplistic non-relativistic two-electron Coulomb-like theory, as well
as the resulting one-hole plus two-hole spectrum.
These predictions correctly describe the apparent features
of the data, but fail here and there by a factor of ${\cal{O}}(2)$.

In Fig.\,(\ref{fig:N1th}b) we ameliorate the data description of Fig.\,(\ref{fig:N1th}a) by
multiplying the prediction for the N1O1up (off) contribution by a factor 2.5 (0.7). Both the
up and off N1N4/5 contributions have been corrected by a factor 1/3, which is the maximum
``tolerated'' by the data, a smaller factor would give a better description, as in Fig.\,(\ref{fig:N1}b).

Even for the single-hole peaks, the theoretical predictions of some of their widths and relative
magnitudes appear to be inaccurate by factors of ${\cal{O}}(2)$. This is not a surprise,
atoms with as many electrons as Ho or Dy are difficult to handle theoretically, in particular
when dealing with their outer orbitals. The real surprise is that our naive treatment of the
two-hole contributions yields results of comparable precision.

The day the data are abundant enough to provide a competitive result on the neutrino
mass, the entire spectrum will have been measured with gigantic statistics. Its various
contributions will have been understood, at least phenomenologically, with the
precision required to distinguish a neutrino mass from any other phenomenon
affecting the spectral shape at the endpoint.

\section{Conclusions and outlook}

Our preliminary analysis of the preliminary low-statistics data yields three main conclusions:
\begin{itemize}
\item{} The shake-off of the electrons in a two-hole process is as relevant as their shake-up.
\item{} The theoretical predictions of the two-hole probabilities do not agree with the data.
\item{} The potentially most important threat to a simple theoretical 
analysis of the end-point of the spectrum
may be the process of M1 capture accompanied by N1 shake-{\it off}.
\end{itemize}

We plan to write a much more detailed paper in which we discuss, along many other relevant points,
the reasons why the current two-hole calculations and the observations may disagree, as well as the
reasons why the last of the above items ought not to be a problem.\\

{\bf Aknowledgements.} ADR acknowledges partial support from the  European Union FP7  ITN INVISIBLES (Marie Curie Actions, PITN- GA-2011- 289442).

%The amplitude for a single N1 electron staying put is:
%\begin{equation}
%A\equiv\langle{\rm Dy[N1^*]}\vert{\rm Dy[N1^*]}\rangle=1-\sum_n B_n^2,
%\end{equation}
%The probability of shake-up of the N1 electron is
%\begin{equation}
%P_{SU}=\sum_{n=7}^\infty |B_n|^2,
%\end{equation}
%and a similar expression holds for the probability of shake-off $P_{SO}$.

\end{document}